\documentclass
[reprint,
]{revtex4-2}

\usepackage{graphicx}
\usepackage{dcolumn}
\usepackage{bm}

\begin{document}

\preprint{APS/123-QED}

\title{Electro magnetic transitions from isobaric analogue states to study nuclear matrix elements for  neutrinoless $\beta \beta $ decays and astro-neutrino inverse $\beta$-decays}  

\author{Hiroyasu Ejiri}
\email{ejiri@rcnp.osaka-u.ac.jp}
 \affiliation{Research Center for Nuclear Physics, Osaka University, Osaka 567-0047, Japan
 }%

\date{\today}
\begin{abstract}

 Experimental studies of nuclear matrix elements (NMEs) for neutrinoless double beta decays (DBDs) and astro-neutrino ($\nu$) inverse beta decays (IBDs) are crucial for $\nu$ studies beyond the standard model and the astro-$\nu$ studies since  accurate theoretical calculations of the NMEs are hard due to  the high sensitivity of the NMEs to the nuclear models and the nuclear parameters used for the models.  The NMEs associated with DBD and 
IBD, including the weak couplings, are found to be experimentally obtained by measuring the corresponding  electromagnetic gamma (EM:$\gamma$) transitions from the isobaric analogue states (IASs) of the DBD and IBD nuclei. Then, the experimental NMEs and the couplings are used for evaluating the DBD and IBD NMEs and for checking the model calculations. The EM-NMEs, the cross-sections and the event rates for the IAS-$\gamma$ transitions are estimated for DBD and IBD nuclei to show the feasibility of the experiments. 
\end{abstract}

\maketitle



{\it Motivation}- Neutrinoless double beta decay (DBD $\beta\beta$), which violates the lepton-number conservation law,  is a sensitive and realistic probe for studying the neutrino ($\nu$) nature (Majorana or Dirac), the absolute $\nu$-mass scale,  the right-handed weak current (RHC) and others, which are  beyond the standard model, as discussed in reviews \cite{eji05,avi08,ver12}. Astro-$\nu$ productions, astro-$\nu$ syntheses, and astro-$\nu$-oscillations are of great interest for physics of astro-neutrinos and supernovae, and are studied by nuclear charged-current (CC) interaction, i.e. inverse beta ($\beta $) decay (IBD). The DBD and IBD rates are proportional to their neutrino nuclear responses, i.e. the squares of the nuclear matrix elements (NMEs).  
The present letter addresses the NMEs, which are indispensable for the $\nu$ studies by these rare nuclear decays. 
 
The DBD $\nu$-mass  process involves the Majorana $\nu$ exchange between two nucleons in the DBD nucleus, where the $\nu$ exchange is much enhanced because of the short distance  between the two nucleons there.  Ton-scale DBD nuclei ($N\approx 10^{28}$ isotopes) are used to access the ultra-rare decay rate  of the order of 10$^{-28}$ per year to study the ultra small $\nu$-mass. Then the accurate value for  the NME for the $\nu$ exchange in the nucleus is required. 
It is, however, extremely hard to calculate the NME by using theoretical models since the NME for the DBD ground state to be studied is a tiny fraction ($\approx$10$^{-4}$) of the total sum of the DBD responses and the NME is very sensitive to all kinds of nuclear correlations  in the DBD nucleus. Thus the calculated NMEs, including the effective axial-vector coupling ($g_{\rm A}^{\rm eff}$), scatter over an order of magnitude, depending on the nuclear models and the effective coupling $g_{\rm A}^{\rm eff}$ and other parameters used in the models \cite{eji00,eji19,fae98,suh12,bar13,eng17,jok18}.
  Thus experimental inputs are crucial to check the theoretical models and the nuclear parameters ($g_{\rm A}^{\rm eff}$ and others) to be used for the calculations \cite{eji00,eji19,jok18,eji20,eji21a,eji22}.  

Astro-$\nu$s such as the solar and supernova $\nu$s are studied via the nuclear CC interaction since the rate is 3-4 orders of magnitude larger than  the electron interaction rate and the ultra-low background experiment is posible by measuring the radioactive decay of the IBD product. Then one needs the astro-$\nu$ NME to study the astro $\nu$ properties and reactions of astro-particle physics interests as discussed in \cite{eji00,eji19}.
Astro-$\nu$ NMEs for the ground and excited states in residual nuclei are required for the astro-$\nu$ studies.  DBD nuclei and some other nuclei are shown to be used for solar-$\nu$ real-time measurements \cite{eji00b,zub03,rag76}. The solar-$\nu$ NMEs for the excited states in $^{71}$Ge are critical in views of the possible $\nu$-oscillation to the sterile $\nu$ \cite{bar22}.  Experimental studies for these astro-$\nu$ NMEs are very valuable  since accurate theoretical calculations for them are hard as for the DBD NMEs  \cite{eji00,eji19}. 

 The present letter aims to show for the first time that the electro-magnetic gamma (EM $\gamma$) NMEs for electric dipole (E1) and magnetic dipole (M1) $\gamma$-transitions from the isobaric analogue states (IASs) of the initial DBD and IBD states are experimentally measured to be used actually for getting the analogous vector dipole (Fermi:V1) and axial-vector dipole (Gamow-Teller:GT) NMEs associated with DBDs and IBDs to help and check the theoretical model calculations for their NMEs. 

Several experimental approaches to the DBD and astro-$\nu$ NMEs  are discussed \cite{eji00,eji19, eji20,eji21a,eji22}. Among them, charge exchange nuclear reactions (CERs) of  ($^3$He,$t$) with sub GeV $^3$He have been used widely to estimate GT NMEs associated with DBD and astro-$\nu$ GT NMEs.  
The CER, however, includes mixed interactions of the GT(spin isospin:$\sigma \tau$), the tensor ([$\sigma \times Y_2$]$_1$), and the vector (isospin:$\tau$) interactions \cite{eji19,hax98,kos19}, and the cross section depends much on the nuclear distortion.   
Thus the CER data provide gross features of the DBD and astro $\nu$-responses, but not exclusively accurate values of the absolute V1 and GT NMEs for individual states relevant to the DBD and astro-$\nu$ NMEs \cite{eji00,eji19,eji20,eji21a,eji22}.  Accurate electron-capture ($\beta ^+$) data are not available for DBD nuclei of current interests due to the extremely small decay-energy, while  $\beta ^-$ decay data are limited to high multipole ones, which are minor components of DBD NMEs. Beta decay data  to be used for the astro-$\nu$ CC NMEs are limited only to the ground state in some nuclei.

Merits of the IAS-EM studies are as follows.
i. EM transition operators are exclusive and well defined ones with the well known EM couplings. Thus the observed E1 and M1 NMEs are used to get the V1 and GT NMEs and effective vector and axial-vector couplings. 
ii. The absolute value for the EM NME is obtained experimentally by measuring  the CER IAS cross section, the IAS width and the CER IAS-$\gamma$ cross section.
iii. The IAS  is a very sharp state, reflecting the isospin symmetry. Thus  backgrounds from non-IAS excitations are small. IAS neutron decays are so reduced that the $\gamma$-branch is enhanced by orders of the magnitude.

One drawback of the EM($\gamma$) experiment is the small cross section due to the small EM interaction in comparison with the strong nuclear interaction for the CERs. It is overcome by using a sharp IAS with the large cross section and the large $\gamma$ branch and also by using large acceptance detectors available recently.


{\it DBD and IBD NMEs}- The DBD rate for the light $\nu$-mass process is expressed as \cite{eji05,ver12,eji19}
\begin{equation}
R^{0\nu}={\rm ln2}~g_{\rm A}^4G^{0\nu} [|m_{\nu} M^{0\nu}|]^2,
\end{equation}
where $G^{0\nu}$ is the phase space factor, $g_{\rm A}$=1.27 is the axial-vector coupling for a free nucleon in units of the vector coupling of $g_V$, $m_{\nu}$ is  the effective $\nu$-mass, and $M^{0\nu}$ is the NME.  $m_{\nu}$ is replaced by the right-handed coupling $\eta $ in case of the RHC process.   $M^{0\nu}$ is given mainly by

\begin{equation}
M^{0\nu}=M^{0\nu}({\rm GT}) + (g_{\rm V}/g_{\rm A})^2M^{0\nu}({\rm V}),
\end{equation}
where $M^{0\nu}$(GT) and $M^{0\nu}$(V) are the axial-vector (GT) and vector (V) NMEs, respectively.   They are expressed in terms of the model NMEs of  
$M^{0\nu}_{\rm GT}$ and  $M^{0\nu}_{\rm  V}$ as  
$M^{0\nu}$(GT)=($g_{\rm A}^{\rm eff}$/$g_{\rm A})^2 M^{0\nu}_{\rm GT}$ and
 $M^{0\nu}({\rm  V})=(g_{\rm V}^{\rm eff}$/$g_{\rm V})^2M^{0\nu}_{\rm  V}$, where $g_{\rm A}^{\rm eff}$ and $g_{\rm V}^{\rm eff}$ are the effective axial-vector and vector couplings introduced to incorporate the renormalization (quenching) effects, i.e. the nuclear many-body and medium effects  that are not included in the model NMEs of $ M^{0\nu}_{\rm GT}$ and $ M^{0\nu}_{\rm V}$\cite{eji19,eji20,eji21a,eji22,suh17,eji19a}. Actually, there is a minor tensor NME in Eq. (2).

The NME $M^{0\nu}(\delta)$ with $\delta$=GT,V is given by the sum of the NMEs $M^{0\nu}_i(\delta)$ for the intermediate states $i$. 
The DBD proceeds as  the double neutron-proton transitions of n$\rightarrow$p and n'$\rightarrow$p' with the $\nu$ exchange between n and n'. Then the GT and V operators  are given by
the double GT (double $\tau\sigma$) and double V (double $\tau$) operators for n$\rightarrow$ p and n'$\rightarrow$p' vial the $\nu$ potential $h(\delta)$ for the virtual-$\nu$ exchange. Then the DBD NME with $h(\delta$) includes the multipole ($J$) components up to around $J\approx$ 6, reflecting the virtual-$\nu$ momentum of $p\approx$100 MeV/c.

The astro-$\nu$ IBD rate is expressed as \cite{eji00,eji19}
\begin{equation}
R^{\nu}={\rm ln2}~g_{\rm A}^2G^{\nu}f_{\nu}|M^{\nu}|^2,
\end{equation}
where $G^{\nu}$ is the phase space factor, $f_{\nu}$ is the $\nu$ flux and  $M^{\nu}$ is the IBD NME. It is given mainly as 
 \begin{equation}
M^{\nu}= M^{\nu}({\rm GT}) + (g_{\rm V}/g_{\rm A}) M^{\nu}({\rm  V}),
\end{equation}
where $M^{\nu}({\rm GT}$) and $M^{\nu}({\rm V}$) are the axial-vector and  vector NMEs for the astro-$\nu$ IBD of $\nu$+n$\rightarrow$e+p.
 They are expressed, as in DBD NMEs, in terms of the model NMEs of  
$M^{\nu}_{\rm GT}$ and  $M^{\nu}_{\rm  V}$ as  
$M^{\nu}$(GT)=($g_{\rm A}^{\rm eff}$/$g_{\rm A}) M^{\nu}_{\rm GT}$ and
 $M^{\nu}({\rm  V})=(g_{\rm V}^{\rm eff}$/$g_{\rm V})M^{\nu}_{\rm  V}$.
Actually, $M^{\nu}({\rm GT})$ is mainly the GT($J=1^+$) NME and $M^{\nu}({\rm  V}$) is the F (Fermi) ($J=0^+$) NME for low energy  $\nu$s $\le$15 MeV.

 $M^{0\nu}$ is sensitive to  the nuclear many-body correlations involved in the double GT and double V transitions of (n$\rightarrow$p, n'$\rightarrow$p') in the DBD process
of $^A_Z$X$\rightarrow ^A_{Z+1}$X$\rightarrow ^A_{Z+2}$X with $A,Z$ being the mass and atomic numbers of the nucleus. The neutron number is $A-Z=N$. Likewise, $M^{\nu}$ is  sensitive to the nuclear many-body correlations involved in the IBD of n$\rightarrow$p in  $^A_Z$X$\rightarrow ^A_{Z+1}$X. 

{\it IAS-EM NMEs for weak NMEs}- The weak NME $M^-(\alpha)$ for the $\alpha$-mode $\beta^-$ transition is related to the EM($\gamma$) NME $M^{IA}(\alpha ')$ for the analogous $\alpha'$ mode IAS-$\gamma$ transition as shown first in  \cite{eji68,eji78} as 
\begin{equation}
M^-(\alpha)\approx \sqrt {2T} M^{\rm IA}(\alpha').
\end{equation} 
where $T=(N-Z)/2$ is the isospin of the initial (ground) state $^A_Z$X, and IAS is expressed as  $(\sqrt{2T})^{-1}$ T$^-$$^A_Z$ X  with T$^-$ being the isospin lowering operator of n$\rightarrow$p. The $\beta ^-$ NMEs $M^{-}(\alpha)$ with $\alpha$=GT and V1 are derived from the analogous IAS-$\gamma$ NMEs $M^{IA}(\alpha ')$ with $\alpha'$=M1 and E1. 
The weak ($\beta $) and $\gamma $ transitions relevant to the DBD, astro-$\nu$ IBD and CER are schematically shown in Fig. 1. 

\begin{figure}[ht]
\hspace{0.2cm}
\vspace{-0cm}
\includegraphics[width=0.44\textwidth]{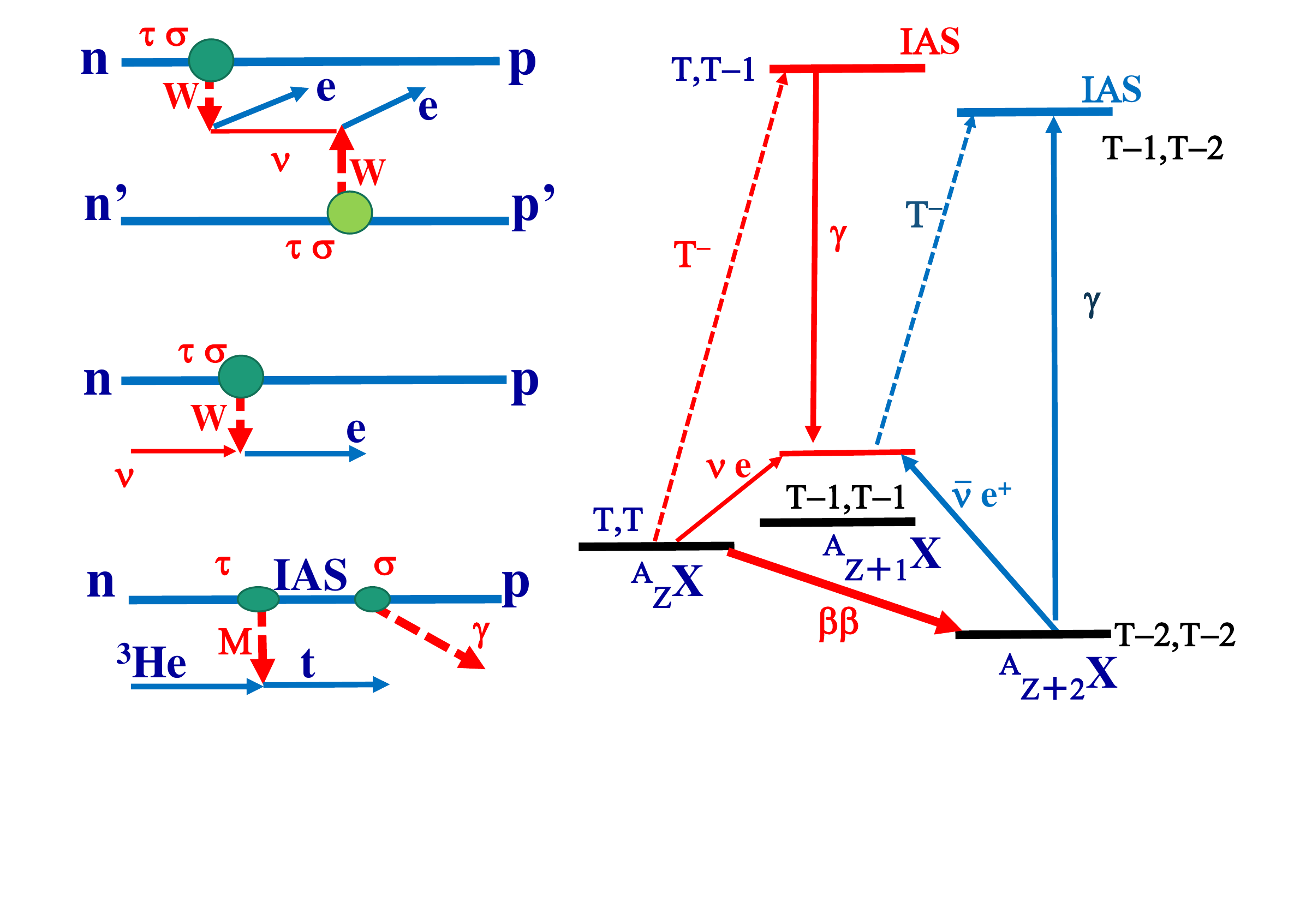}
\vspace{-1.2cm}
\caption{Left side: GT weak and IAS-M1$\gamma$ transition diagrams. Top: DBD with $\nu$ exchange between n and n'. Middle: astro-$\nu$ IBD. Bottom: ($^3$He,$t$) CER of the  IAS excitation followed by EM($\gamma$) transition. W: weak boson. M: meson.
 $\tau$: isospin. $\sigma$: spin. Green circles: nuclear $\tau\sigma$ vertexes. Green ellipsoids: nuclear CER ($\tau$) and  M1($\sigma$) vertexes. In case of E1 the operator $\sigma$ is replaced by $rY_1$. \\
Right side: DBD scheme for $^A_{Z}$X$\rightarrow ^A_{Z+2}$X via the intermediate nucleus $^A_{\rm Z+1}$X and the IAS-$\gamma$ transition scheme for the $M^-$ side (red lines) and that for the $M^+$ side (blue lines). ($T,T_z$) are the isospin and it's z component. 
\label{figure:fig1}} 
\end{figure}

In fact, IAS-$\gamma$s for the first forbidden $\beta $ NMEs were studied before by using (p.$\gamma$) reactions via the strong IAS resonance with the large cross section around 10$\mu$b \cite{eji68}. However, there are no stable target nuclei for the (p,$\gamma$) reactions on DBD and astro $\nu$ nuclei.

The IAS-$\gamma $ study for axial vector and vector NMEs is found to be made by using the medium energy ($^3$He,$t$) CER with $E$($^3$He)$\approx$0.42-0.45 GeV. Actually the CER excites strongly the IAS and GT states, and  is used to study GT states in DBD and astro-$\nu$ nuclei  \cite{aki97,eji98,fre11,thi12,thi12b,pup11,gue11,thi12a,pup12,fre16,eji16a,aki20,fre13}. 

IAS is the isospin ($\tau^-$) giant resonance (GR), which is strongly excited as a sharp peak in the $\tau^-$ CER.  The excitation spectrum for a DBD nucleus of $^{82}$Se \cite{fre16} is shown as an example in Fig.2.

\begin{figure}[htb]
\hspace{-0cm}
\vspace{-1cm}
\includegraphics[width=0.43\textwidth]{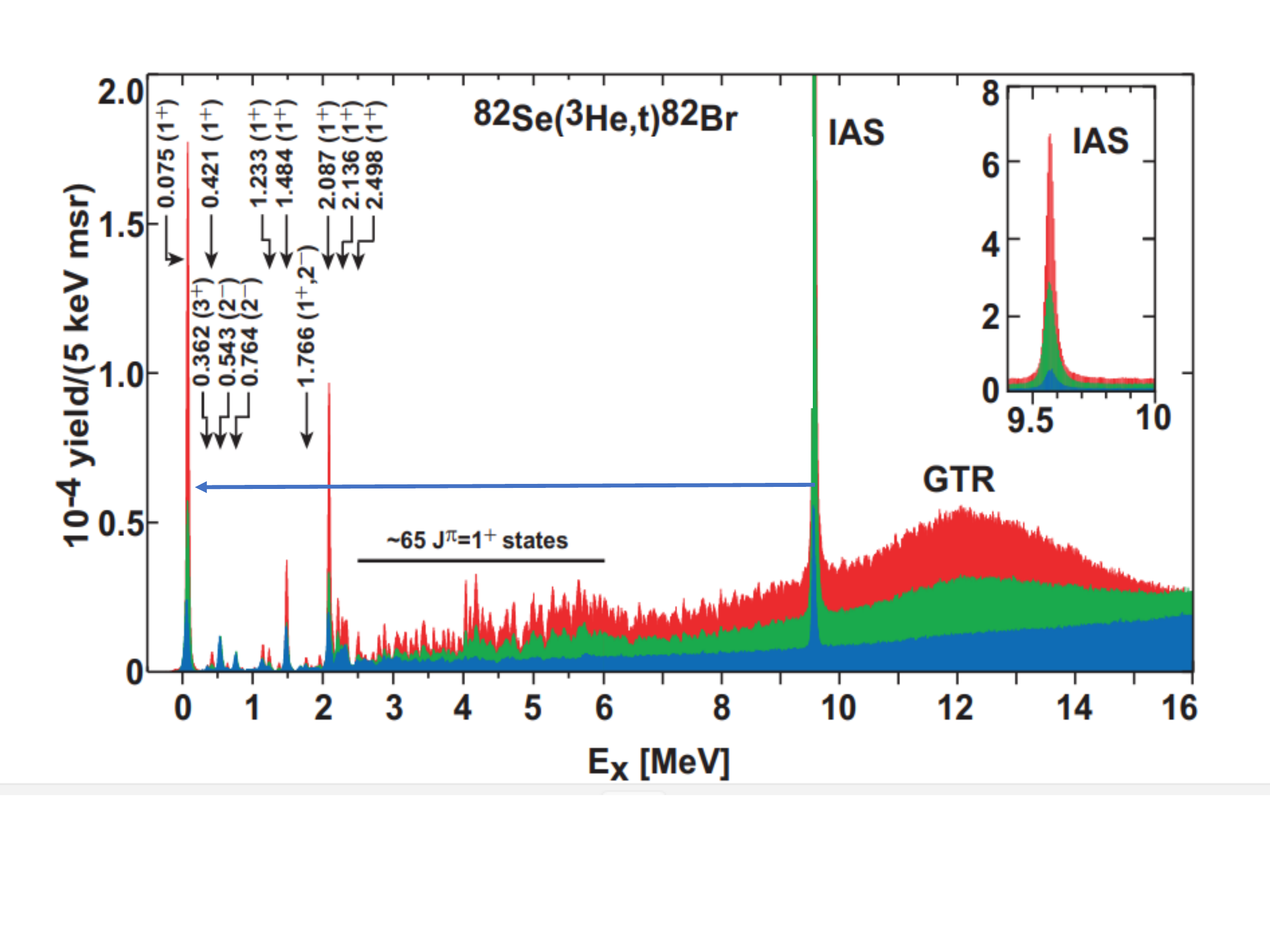}
\vspace{-0.2cm}
\caption{ The energy spectrum of the  ($^3$He,$t$) reaction on $^{82}$Se \cite{fre16}.  The energy scale below 6 MeV is enlarged.  Yields at the $t$ emission-angles of 0-0.5, 1-1.5, and 2-2.5, all in deg., are shown by red, green and blue, respectively. IAS and GT CERs with no angular-momentum transfer are enhanced at the forward angle (red). The blue arrow is $\gamma$ from IAS.
\label{figure:fig3}} 
\end{figure}

The IAS differential cross section is expressed as 
\begin{equation}
d\sigma^{\rm IA}/d\Omega=kNJ_{\tau}^2B({\rm IAS}), 
\end{equation}
with $k, N, J_{\tau}$, and  $B$(IAS) being the kinematical factor including the small effect of the momentum transfer, the distortion factor, the interaction strength and the IAS reduced width. $B$(IAS) is given by the sum rule limit of $2T_z=(N-Z)$. Thus the $d\sigma^{\rm IA}/d\Omega$ at 0 deg. for the DBD nuclei with $A$=70-160 gets as large as 10 mb/sr. 

 The IAS-$\gamma$ differential cross section is measured in coincidence with the IAS CER. It is given as,  
\begin{equation}
\frac{d\sigma^{\rm IA}(\alpha')}{d\Omega}=\frac{d\sigma^{\rm IA}}{d\Omega} \frac{\it \Gamma ^{\rm IA}(\alpha')}{\it \Gamma (\rm T)},
\end{equation}
where ${\it \Gamma ^{\rm IA}(\alpha')}/{\it \Gamma (\rm T)}$ is  the $\gamma$-branching ratio with ${\it \Gamma}^{\rm IA}({\alpha '}$) being the $\alpha '$ mode EM $\gamma$ width and  ${\it \Gamma}$(\rm T) being the total IAS width. Hereafter we discuss mainly IAS-M1 and IAS-E1 ($\alpha '$ =M1,E1) transitions to axial-vector GT(1$^+$) and vector dipole V1(1$^-$) states.

The IAS-$\gamma $ width  ${\it \Gamma}^{\rm IA}({\alpha'})$ in units of eV for $\alpha '$=M1, E1 is expressed in terms of the $\gamma $ ray energy $E_{\alpha'}$ in units of MeV and the $\gamma$ reduced width $B^{\rm IA}(\alpha ')$ as \cite{boh75,eji89}
\begin{equation}
{\it \Gamma}^{\rm IA}(\alpha') = K_{\alpha'}E_{\alpha'}^3 B^{\rm IA}(\alpha'),
\end{equation}
where $K_{\alpha'}$ is the kinematical factor. The reduced width is expressed by using the IAS-EM NME $M^{\rm IA}(\alpha')$ as 
\begin{equation}
B^{\rm IA}(\alpha')=g_{\alpha'}^2|M^{\rm IA}(\alpha')|^2 S^{-1}, 
\end{equation}
where $S=2J_i+1$ is the spin factor with $J_i$ being the initial state spin, and $g_{\alpha'}$ is the 
 EM coupling. The coupling is $g_{M1}= (e\hbar/2Mc) g$ with $e, M, c, g$ being the electron charge, the nucleon mass, the light velocity, and  M1 $g$ factor  for the M1 $\gamma$, and  $g_{\rm E1}$= e  for the E1 $\gamma$. $M$(E1) is in units of fm.
The kinematical factor is 1.05 for E1 and 1.16 10$^{-2}$ for M1 transitions.

The M1 and E1 $\gamma$ NMEs are related to the corresponding GT and V1 $\beta$ NMEs.  In fact, M1 transition operator includes a  contribution from the orbital term in addition to the spin term. In case of a spin-streched transition of $l\pm{1/2}\leftrightarrow l\mp{1/2}$, the M1 $g$ factor is given as $g=\sqrt{3/4\pi}
(g_s/2-g_l/2$) with $g_s$ and $g_l$ being the nucleon spin and orbital $g$ factors. Here note $g_s\gg g_l$. Actually, the major GT transitions are the spin-streched transitions of g$_{7/2}\rightarrow$g$_{9/2}$ and d$_{3/2}\rightarrow$d$_{5/2}$ for the present DBD nuclei with $A\approx$96-116 and $A\approx$128-136, respectively.
Then the  MI$\gamma$-NME $M^{\rm IA}({\rm M1})$ with the transition operator of T(M1)=$\sigma$ is used to get the analogous GT$\beta$-NME with the transition operator of  T(GT)=$\tau \sigma$.

In case of the E1 transition, the transition operator is T(E1)=$rY_1$ with $r$ and $Y_1$ being the radius and the spherical Bessel function. Then EI$\gamma$-NME $M^{\rm IA}({\rm E1})$ with the transition operator of T(E1) is used to get the analogous V1$\beta$-NME with the transition operator of  T(V1)=$\tau rY_1$.  

Actual procedures to get the $\beta$ NMEs associated with the DBD and astro-$\nu$ IBD NMEs are  i. excite the IAS by the ($^3$He,t) reaction on $^A_Z$X (see Fig.1) and measure the IAS differential cross section $d\sigma^{IA}/d\Omega$ and the IAS total width ${\it \Gamma}$(T), ii. obtain the IAS-$\gamma$ differential cross section $d\sigma^{IA}$($\alpha'$)/$d\Omega$ with $\alpha'$=M1, E1 for low-lying 1$^+$ and 1$^-$ states in $^A_{Z+1}$X by measuring the $\gamma$-rays in coincidence with the  IAS CER,  iii. obtain the reduced $\gamma$ widths ${\it \Gamma}^{\rm IA}$($\alpha'$) by using the measured IAS and IAS-$\gamma$ differential cross-sections  and the measured total width ${\it \Gamma}$(T) in eq.(7), iv. obtain the reduced $\gamma $  width $B(\alpha')$ from the measured $\gamma$ width ${\it \Gamma} (\alpha')$ in eq. (8), and the $\gamma$ NME $M^{\rm IA}$ ($\alpha'$) by using the $B(\alpha'$)  in eq.(9), and  v. get the $\beta ^-$ NMEs $M^-$($\alpha$) with $\alpha$=GT,V1 from the  $M^{\rm IA}(\alpha'$) with $\alpha'$=M1,E1 in eq.(5).  

{\it IAS-$\gamma$ cross sections and event rates}- The IAS-$\gamma $ cross sections for realistic cases of the GT and V1 T$^-$ transitions and the experimental counting rates for them are estimated  to  show how these measurements are feasible. 

The IAS-M1 widths ${\it \Gamma}_{\rm M1}$ for  GT states in DBD and IBD nuclei are estimated by using the NMEs $M$(GT)  for  low-lying states in the DBD and astro-$\nu$ nuclei.  
We use the GT states with relatively large $B$(GT) of the order of 10$^{-1}$, which are considered to be the spin-stretched GT transitions.   The reduced widths $B$(M1) are estimated by assuming  the CER $B$(GT) derived without considering the tensor term interference \cite{aki97,thi12,thi12b,pup11,gue11,thi12a,pup12,fre16,eji16a,aki20,fre13}.  The IAS-M1 $\gamma $ widths ${\it \Gamma}_{\rm M1}$ are estimated as shown in Table 1.  

 The IAS-E1 $\gamma $ widths ${\it \Gamma}_{\rm E1}$ for  V1 1$^-$ states in DBD nuclei are estimated by using the QP (quasi-particle) model V1 NMEs $M({\rm V}1)$  since the V1 $\beta$ reduced widths in DBD  nuclei are not known experimentally. The V1 NMEs for QP states in medium heavy nuclei have been shown to be approximatrly given  by the QP model with experimental effective coupling of $g_{\rm E1}^{\rm eff}$/$g_{\rm E1}\approx$ 0.2 - 0.25 \cite{eji68,eji78,eji89}.  So the V1 NME for the typical V1 transition of n(1h$_{11/2}$)$\rightarrow$p(1g$_{9/2}$) in  3 DBD nuclei are estimated by using the QP NMEs with the effective coupling of $g_{\rm E1}^{\rm eff}$/$g_{\rm E1}$=0.225. They are shown in Table 2. 

The IAS-$\gamma $ branching ratio ${\it \Gamma}^{\rm IA}({\alpha'}$)/${\it \Gamma}$(T) is obtained by using the estimated ${\it \Gamma}_{\alpha'}$ and the total width ${\it \Gamma }$(T) measured experimentally. The total widths for the DBD nuclei are derived from the high energy-resolution CERs \cite{aki97,thi12,thi12b,pup11,gue11,thi12a,pup12,fre16,eji16a,aki20,fre13} as shown in Fig.3, together with other experimental widths  \cite{har86}. They are given by 
\begin{equation}
{\it \Gamma }_T\approx 3.5 T_z  ~{\rm keV},~~~T_z=(N-Z)/2.
\end{equation}
The IAS width is indeed very small in comparison with a typical neutron width of the order of MeV. 

Then, the IAS-M1$\gamma$ and IAS-E1$\gamma$ cross sections at 0 deg. are estimated by using the measured IAS cross section at 0 deg., the experimental ${\it \Gamma} $(T) and the
estimated $B$(M1) and $B$(E1) values as given in Tables 1 and 2. The IAS-M1$\gamma$ and IAS-E1$\gamma$ cross sections are of the orders of 10-100 nb and 100-1000 nb/sr, respectively.
\begin{figure}[htb]
\hspace{0cm}
\vspace{0.2cm}
\includegraphics[width=0.3\textwidth]{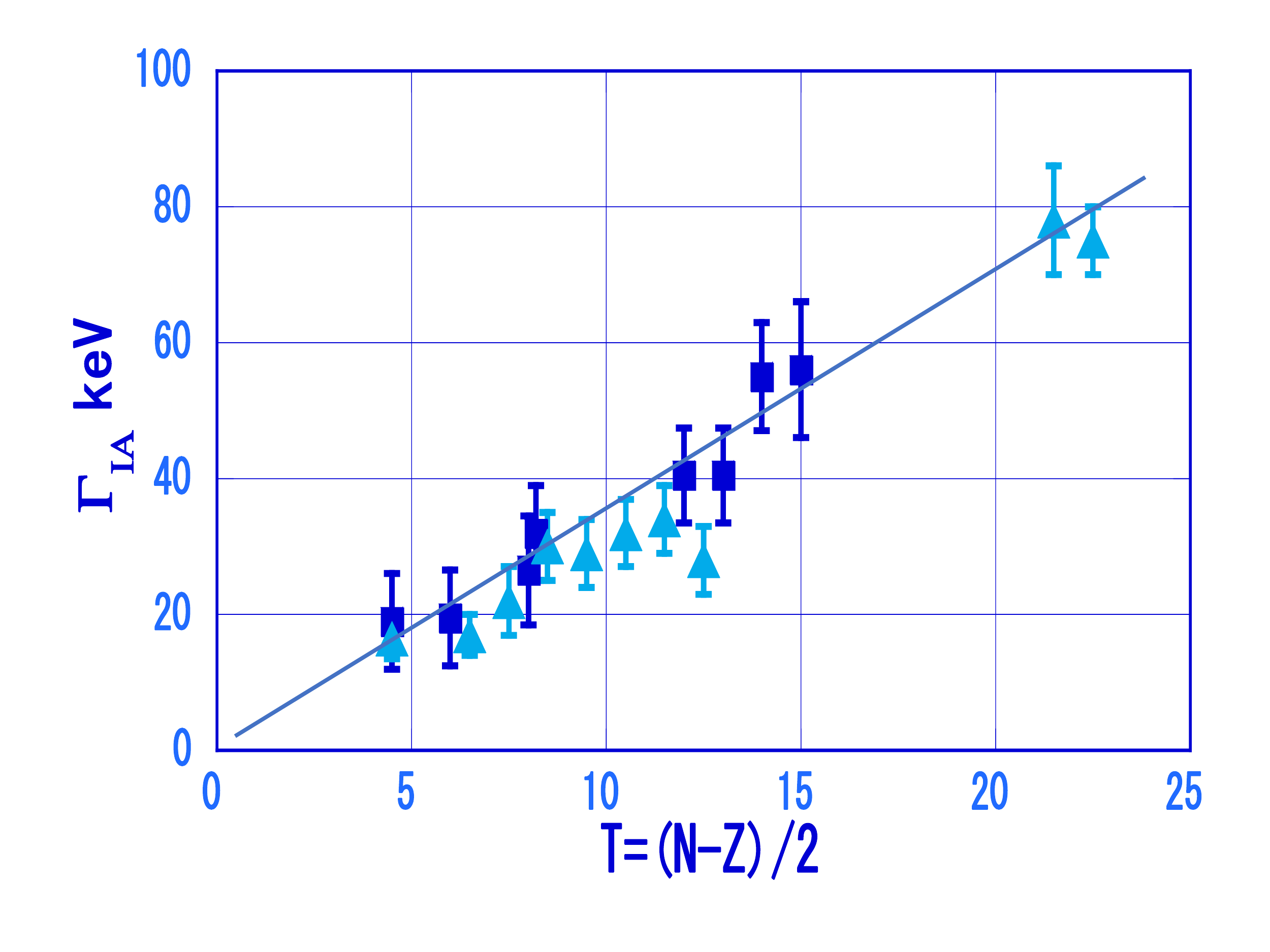}
\vspace{-0.8cm}
\caption{Experimental IAS widths as a function of the isospin $T_z=(N-Z)/2$. Blue sqaures for IASs in DBD nuclei derived in the present work. Light blue triangles from ref. \cite{har86}.
\label{figure:fig3}} 
\end{figure}
\begin{table}[htb] 
\caption{M1$\gamma $ widths and the IAS $\gamma $ cross sections estimated for DBD nuclei of current interests and $^{71}$Ga for the solar $\nu$s.
Shown are  $E$(IAS) and $E$(GT) in units of MeV, $B$(GT), $B$(M1) in units of 10$^{-2}$, ${\it \Gamma} $(M1) in units of 10$^{-2}$ eV, and  $\sigma$(M1)=d$\sigma^{\rm IA}$(M1)/$d\Omega$ in units of nb (10$^{-9}$b)/str.
\vspace{0.2cm}}
\centering
\begin{tabular}{ccccccc}
\hline
$A$ & $E$(IAS) & $E$(GT)  & $B$(GT) & $B$(M1)  & ${\it \Gamma} $(M1)  & $\sigma$(M1) \\
\hline
$^{76}$Ge~ &8.31 & 1.07 & 0.14  & 1.45 & 6.4&41\\
$^{82}$Se~ &9.58 & 0.075 & 0.34  & 3.0 & 30.0&150\\
$^{96}$Zr~  &10.9  & 0.69& 0.16  & 1.25 &15.3&76\\
$^{100}$Mo~ &11.1 & 0   &0.35 &  2.7 & 43.4& 170  \\
$^{116}$Cd~ &12.1 & 0   &0.14 &  0.88 & 18.0& 51  \\
$^{128}$Te~ &12.0 & 0 & 0.079 & 0.41  & 8.2  &17\\
$^{130}$Te~ &12.7 & 0 & 0.072& 0.35 & 8.2  &17\\
$^{136}$Xe~ &13.4 & 0.59 &0.23 & 1.03  & 25  &45\\
$^{150}$Nd~&14.4& 0.11 & 0.13 &0.54 & 18.0 &35\\
\\
$^{71}$Ga~ &8.91 & 0 & 0.085  & 1.2 & 9.8 &51 \\

\hline
\end{tabular}
\end{table}

\begin{table}[htb] 
\caption{E1$\gamma $ widths and the IA$\gamma $ cross sections estimated for 3 DBD nuclei.
Shown are  $E$(IAS) and $E$(V1) in units of MeV, QP model $B$(V1), $B$(E1) in units of 10$^{-2}$, ${\it \Gamma}$(E1) in units of  10$^{-2}$eV and  
$\sigma$(E1)=d$\sigma^{\rm IA}$(E1)/d$\Omega$ in units of nb (10$^{-9}$b)/str.
\vspace{0.25cm}}
\centering
\begin{tabular}{ccccccc}
\hline
$A$ & $E$(IAS) & $E$(V1)  & $B$(V1) & $B$(E1)  & ${\it \Gamma}$(E1)  & $\sigma$(E1) \\
\hline

$^{96}$Zr~  &10.9  & 3& 6.8  & 43 &220 &1080\\
$^{100}$Mo~ &11.1 & 3   &7.5 &  47 & 260& 1020  \\
$^{130}$Te~ &12.7 & 3& 1.0 & 3.8  & 36 &75\\

\hline
\vspace{0cm}
\end{tabular}
\end{table}

\begin{figure}[htb]
\hspace{0cm}
\vspace{-0.cm}
\includegraphics[width=0.43\textwidth]{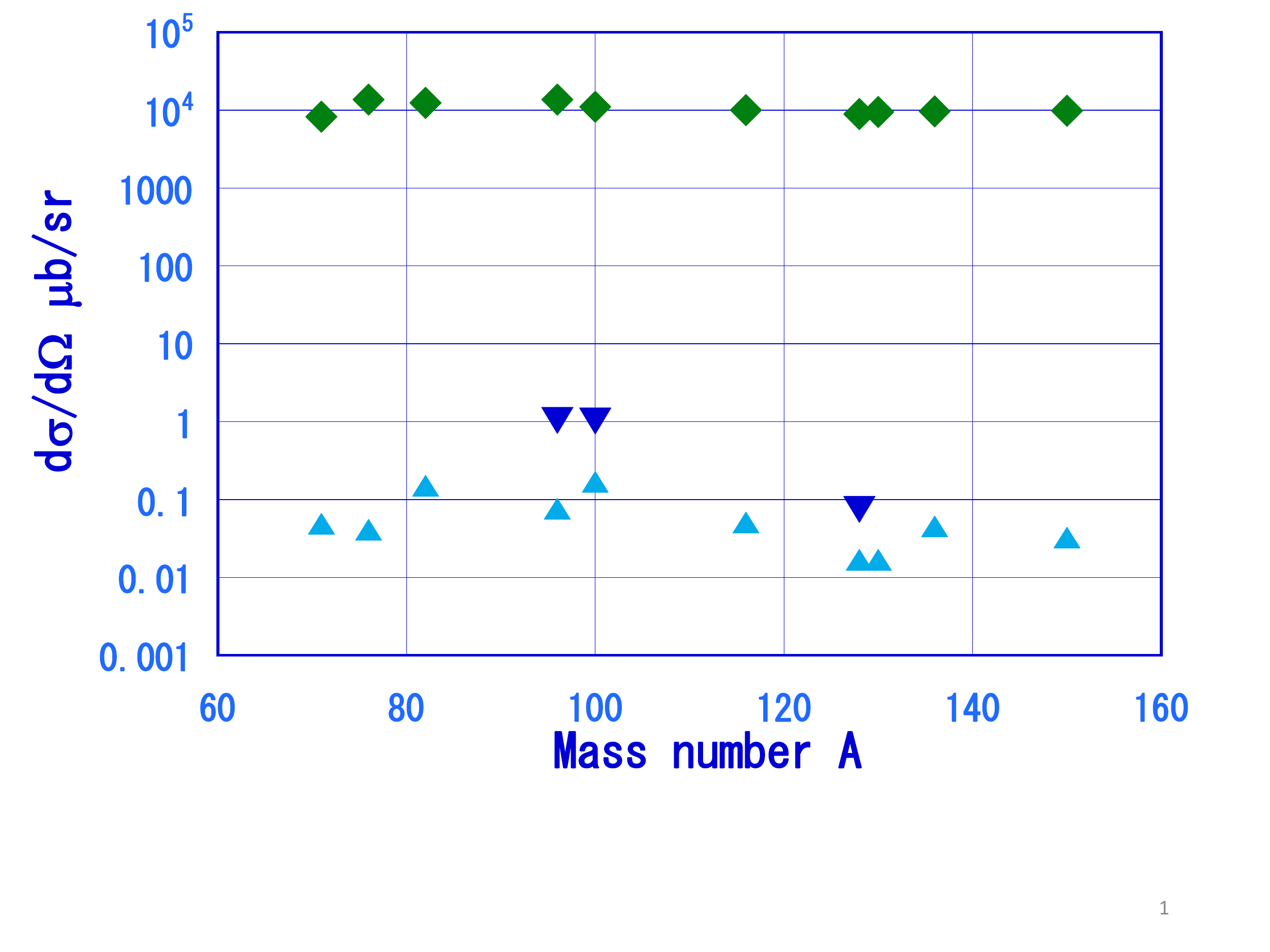}
\vspace{-1.3 cm}
\caption{Differential cross sections at 0 deg. for  DBD and astro-$\nu$ nuclei.  Green diamonds: IAS excitations. Blue: inverse triangles: IAS-$E1\gamma$s to 1$^-$ states. Light blue triangles: IAS-M1$\gamma$s to 1$^+$ states.   
\label{figure:fig1}} 
\end{figure}
The estimated IAS-$\gamma $ cross sections for the M1 and E1 transitions are about 10$^{-6}$-10$^{-4}$ of the IAS ones of 10 mb, reflecting the M1 and E1 $\gamma$ branching ratios to the total IAS width. The E1 cross sections are larger by 1-2 orders of magnitude than the M1 cross sections because of the larger kinematical factor in eq.(8). The very small $d^{\rm IA}\sigma({\rm E1})/d\Omega$ for $^{130}$Te is due to the small vacancy factor $U$ for the 1g9/2 proton. 

 IAS-$\gamma $ event rates under typical experimental conditions are estimated to show  feasibility of the experimental studies.  Using a target of 40 mg/cm$^2$, a $^3$He beam of 20 npA, and a spectrometer of a solid angle of 3.2 msr,  the IAS-$\gamma $ event rate per day is
\begin{equation}
R_{\gamma}=8A^{-1} 10^2 (d\sigma^{\rm IA}(\alpha')/d{\Omega})\epsilon_{\gamma}, 
\end{equation}
where $A$, $d\sigma^{\rm IA}({\alpha'})$/$d{\Omega}$, and $\epsilon_{\gamma}$ are the target mass number, the differential cross section with $\alpha'$=M1,E1 in units of nb/sr, and the $\gamma$ detection efficiency. Recent developments of  $\gamma$ -detectors have made it possible to measure $\gamma$-rays with the efficiency around 5-10 $\%$ \cite{gia13,kob19,tam22}. Then on gets $Y\approx$50 per day in case of $\epsilon \approx$6$\%$ and d$\sigma^{\rm IA}(\alpha'$)/$d\Omega \approx$100 nb. Then  the IAS-$\gamma$ experiments to get accurate ($\pm$10$\%$ or so) NMEs are quite feasible. 

The IAS cross section (eq.(6)) remains almost same in a wide region of the projectile energy $E$ since the interaction $J_{\tau}$ decreases with  $E$ \cite{fra85}, but the kinematical factor $k$ increases as $E$. Thus one may use any medium-energy  accelerators for the experiments. 

{\it Concluding remarks }-  The IAS-EM($\gamma$) study opens a new experimental way to access the  weak NMEs for the DBD and astro-$\nu$ NMEs, which are crucial for neutrino studies in complex nuclei.  It provides the absolute values for  M1-E1 NMEs with well-defined EM operators and the EM couplings, which  are used to obtain the corresponding weak GT-V1 NMEs associated with DBD NMEs and astro-$\nu$ IBD NMEs. They are free from uncertainties due to the tensor and other interfering NMEs and the distortion potential as in the nuclear CERs  used so far. The  IAS-$\gamma $ coincidence measurement leads to background-free measurements.  The event rates estimated for realistic GT and V1 states in DBD and astro-$\nu$ nuclei indicate that the experiment is quite realistic and feasible  with  available detectors.  
 
Then  DBD and IBD nuclear models and their nuclear parameters ($g_{\rm A}^{\rm eff}$, $g_{\rm V}^{\rm eff}$, $g_{pp}$, etc)  are well checked by comparing the  GT and V1 NMEs (and also IAS-M1 and IAS-E1 NMEs) calculated by using the same DBD and IBD models with the experimental GT and V1 NMEs (and also IAS-M1 and IAS-E1 NMEs) derived from the IAS-EM$\gamma$ experiments. The experimental $g_{\rm M1}^{\rm eff}$ and $g_{\rm E1}^{\rm eff}$ help evaluate the $g_{\rm A}^{eff}$ and $g_{\rm V}^{eff}$ used for DBD and IBD  models since calculations for them are  hard. 

The experimental GT and V1 NMEs studied by the IAS-EM($\gamma $),  together with the experimental $\beta$ and 2$\nu\beta \beta $ NMEs and the experimental GT and SD (spin-dipole) GRs studied by CERs \cite{eji22}, are very powerful in pinning down the DBD and astro-$\nu$ NMEs.



EM($\gamma$) excitations of IAS, which are inverse reactions of the present $\gamma $ decays from IAS, are used for studying $M^+(\alpha$) NMEs associated with DBDs and astro-$\bar{\nu}$s as shown in Fig. 1 and in \cite{eji13}.

Studies of EM($\gamma$) decays from GT-GR excited by CER provide the GT NME and the $g_{\rm A}^{\rm eff}$ for the GT GR.  Comparison of GT NMEs derived from the IAS-M1 ($\gamma$) studies and the GT NMEs derived from CERs is interesting to see the possible tensor interference effect. 

The present single CER IAS-$\gamma$ emphasizes accurate and exclusive measurements of the NMEs associated with the DBD and astro-$\nu$ NMEs. Double CERs and double $\gamma$s are also interesting challenges in future \cite{cap23,rom22}.

\end{document}